\documentclass[11pt]{article}
\usepackage{amssymb}

\newcommand{\bea}{\begin{eqnarray*}}
\newcommand{\eea}{\end{eqnarray*}}
\newcommand{\alfa}{\alpha}

\newcommand{\fie}{\phi}

\input{Qcircuit.tex}
\title{The block-ZXZ synthesis \\ of an arbitrary quantum circuit}
\author{Alexis De Vos$^1$ and Stijn De Baerdemacker$^{2,3}$ \\[2mm]
           $^1$ Cmst, Imec v.z.w. \\
           vakgroep elektronica en informatiesystemen, \\
           Universiteit Gent, B - 9000 Gent, Belgium \\
           $^2$ Center for Molecular Modeling, \\
           vakgroep fysica en sterrenkunde, \\
           Universiteit Gent, B - 9000 Gent, Belgium \\
           $^3$ Ghent Quantum Chemistry Group, \\
           vakgroep anorganische en fysische chemie, \\
           Universiteit Gent, B - 9000 Gent, Belgium}

\begin{document}

\maketitle

\begin{abstract}
Given an arbitrary $2^w \times 2^w$ unitary matrix~$U$,
a powerful matrix decomposition can be applied,
leading to four different syntheses of a $w$-qubit quantum circuit
performing the unitary transformation.
The demonstration is based on a recent theorem
by F\"uhr and Rzeszotnik,
generalizing the scaling of single-bit unitary gates ($w=1$)
to gates with arbitrary value of~$w$.
The synthesized circuit consists of
controlled 1-qubit gates, such as
{\tt NEGATOR} gates and
{\tt  PHASOR} gates.
Interestingly,
the approach reduces to a known synthesis method
for classical logic circuits consisting of controlled
{\tt  NOT} gates,
in the case that $U$ is a permutation matrix.
\end{abstract}

\section{Introduction}

The group U($2^w$), i.e.\ the group of $2^w \times 2^w$ unitary matrices, 
describes all quantum circuits acting on $w$~qubits \cite{nielsen}.  
In the literature, many different decompositions of a unitary matrix~$U$ 
have been proposed to synthesize quantum circuits performing the transformation~$U$.  
These decompositions can be classified into two categories.  
The first category of decompositions reduces the dimension of the unitary matrix with one unit, 
leading to a matrix sequence U($n$), U($n-1$), U($n-2$),\dots. all the way down to U(2).  
Notable examples are based on beam-splitter transformations \cite{reck} or 
the Householder decompositions \cite{ivanov,urias,cabrera}.  
Although these decompositions can be realized physically 
by means of multi-beam splitters or Mach-Zehnder interferometers \cite{reck}, 
they are not in natural accordance with a multi-qubit architecture.  
For this, the second category of decompositions is better suited, 
to which the cosine-sine (CSD) \cite{mottonen}, 
Cartan's KAK \cite{khaneja,bullock}, Clifford-$T$ \cite{bocharov,soeken}, 
and related decompositions \cite{barenco,shende} belong.  
This category reduces a unitary transformation on $w$ qubits, or $w$-qubit gate, 
to a cascade of unitary transformations on $(w-1)$-qubits.  

Recently, it was demonstrated \cite{gent}, in the framework of the ZXZ matrix decomposition, 
that two subgroups of U($n$)
are helpful:
\begin{itemize}
\item XU($n$), the group of $n \times n$          unitary matrices with all line sums    equal to~1;
\item ZU($n$), the group of $n \times n$ diagonal unitary matrices with upper-left entry equal to~1.
\end{itemize}
They allow the implementation of quantum circuits \cite{freiberg}, 
with the help of $2 \times 2$ {\tt PHASOR} gates and $j \times j$ Fourier-transform gates 
with $2 \le j \le 2^w$, which can be realised respectively as phase shifters 
and as $2n$-multiports in $n$-mode quantum-optical circuits \cite{reck,mattle,idel}. 
However compact and elegant in mathematical form, 
the ZXZ decomposition belongs to the first category of decompositions, 
and is not naturally tailored to qubit-based quantum circuits.  
This is due to the presence of the $j \times j$ Fourier transforms, 
which act on a $j$-dimensional subspace of the total $2^w$ Hilbert space, 
rather than on a subset of the $w$~qubits.  
The reason for this is the decomposition of an arbitrary XU($j$)~matrix as
\[
F_j\ \left( \begin{array}{cc} 1 & \\ & U \end{array} \right) F_j \ ,
\]
where $F_j$ is the $j \times j$ Fourier matrix and $U$ is an appropriate U($j-1$)~matrix. 
Hence, the size of the matrix to be synthesized lowers only one unit:
from~$j$ to~$j-1$.  

Below we will demonstrate that a similar but more natural ZXZ-inspired method exists 
which respects the qubit structure of the quantum circuit to be synthesized.   
At each step, the size of the unitary matrix is reduced by a factor~1/2, 
so instead of a matrix sequence from U($n$), U($n-1$), U($n-2$),\dots we will 
take matrices from U($n$), U($n/2$), U($n/4$),\dots.  
On the one hand, this means that the method is not applicable for arbitrary~$n$, 
but only useful for $n$ equal to some power of~2, i.e.\ for $n=2^w$.   
On the other hand, the decomposition is more in line with classical reversible decompositions, 
respecting the bit-structure of the architecture \cite{devos}.  
Indeed, we will also prove that the proposed block-ZXZ decomposition 
leads to the Birkhoff decomposition of classical reversible circuits 
when the unitary matrix is a permutation matrix, 
in contrast to previously proposed methods 
\cite{mottonen,khaneja,bullock,bocharov,soeken,barenco,shende}.

\section{Circuit decomposition}

De Vos and De Baerdemacker \cite{gent,scaling} noticed
the following decomposition of an arbitrary member~$U$ of U(2):
\begin{equation}
U = \left( \begin{array}{cc} a & 0 \\ 0 & b  \end{array} \right)
    \frac{1}{2}
    \left( \begin{array}{cc} 1 + c & 1 - c \\ 1 - c & 1 + c  \end{array} \right)
    \left( \begin{array}{cc} 1 & 0 \\ 0 & d  \end{array} \right) \ ,
\label{U2}
\end{equation}
where $a$, $b$, $c$, and $d$ are complex numbers with unit modulus.
Idel and Wolf \cite{idel} proved a generalization,
conjectured in \cite{scaling}, for
an arbitrary element~$U$ of U($n$) with arbitrary~$n$:
\[
U = Z_1 X Z_2 \ ,
\]
where
$Z_1$ is an $n \times n$ diagonal unitary matrix,
$X$   is an $n \times n$          unitary matrix with all line sums equal to~1, and
$Z_2$ is an $n \times n$ diagonal unitary matrix with upper-left entry equal to~1.
F\"uhr and Rzeszotnik \cite{fuhr} proved an other generalization for
an arbitrary element~$U$ of U($n$), however restricted to even $n$~values:
\begin{equation}
U = \left( \begin{array}{cc} A & 0 \\ 0 & B  \end{array} \right)
    \frac{1}{2}
    \left( \begin{array}{cc} I + C & I - C \\ I - C & I + C  \end{array} \right)
    \left( \begin{array}{cc} I & 0 \\ 0 & D  \end{array} \right) \ ,
\label{F&R}
\end{equation}
where $A$, $B$, $C$, and $D$ are matrices from U($n/2$)
and $I$ is the $n/2 \times n/2$ unit matrix.
We note that, in both generalizations,
the number of degrees of freedom is the same in the lhs and rhs of the equation.
In the former case we have
\[
n^2 = n + (n-1)^2 + (n-1) \ ;
\]
in the latter case we have
\[
n^2 = 2 \left( \frac{n}{2} \right)^2 
      + \left( \frac{n}{2} \right)^2 + \left( \frac{n}{2} \right)^2 \ .
\]

If $n$ equals $2^w$, then the decomposition (\ref{F&R}) allows a circuit interpretation.
Indeed, we can write
\[
  \left( \begin{array}{cc} I + C & I - C \\ I - C & I + C  \end{array} \right) =
F \left( \begin{array}{cc} I     &        \\      &     C  \end{array} \right) F^{-1} \ ,
\]
where $F$ is the following $n \times n$ complex Hadamard matrix \cite{tadej}:
\[
F = \frac{1}{\sqrt{2}}\ \left( \begin{array}{cc} I & I \\ I & -I  \end{array} \right) \
  = H \otimes I \ ,
\]
with $I$ being again the $n/2 \times n/2$ unit     matrix, 
and  $H$             the $2   \times   2$ Hadamard matrix.
We conclude that an arbitrary quantum circuit acting on $w$~qubits
can be decomposed into two Hadamard gates
and four quantum circuits acting on $w-1$~qubits and controlled by the remaining qubit:
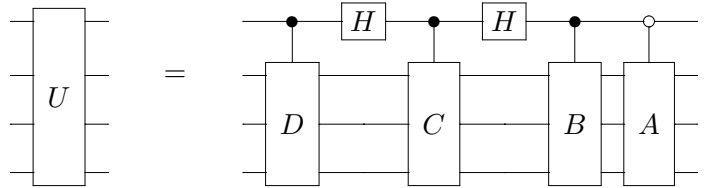
\[
\Qcircuit @C=3mm @R=3mm {
  & \multigate{3}{U} & \qw &&&   &&& & \ctrl{1}         & \gate{H} & \ctrl{1}         & \gate{H} & \ctrl{1}         & \ctrlo{1}        & \qw \\
  & \ghost{U}        & \qw &&& = &&& & \multigate{2}{D} & \qw      & \multigate{2}{C} & \qw      & \multigate{2}{B} & \multigate{2}{A} & \qw \\
  & \ghost{U}        & \qw &&&   &&& & \ghost{D}        & \qw      & \ghost{C}        & \qw      & \ghost{B}        & \ghost{A}        & \qw \\
  & \ghost{U}        & \qw &&&   &&& & \ghost{D}        & \qw      & \ghost{C}        & \qw      & \ghost{B}        & \ghost{A}        & \qw \hspace{4mm}  \ .
}
\]

We now can apply the above decomposition
to each of the four circuits $A$, $B$, $C$, and~$D$.
By acting so again and again,
we finally obtain a decomposition into
\begin{itemize}
\item $h=2(4^{w-1}-1)/3$~Hadamard         gates, and
\item $g=  4^{w-1}$  non-Hadamard quantum gates acting on a single qubit.
\end{itemize}
As the former gates have no parameter and
each of the latter gates has four parameters, the circuit has $4g=4^w$ parameters,
in accordance with the $n^2$ degrees of freedom of the matrix~$U$.
We note that all $h+g$ single-qubit gates
are controlled gates, with the exception of two Hadamard gates
on the first qubit.

One might continue the decomposition
by decomposing each single-qubit circuit into exclusively
{\tt NEGATOR} gates and {\tt PHASOR} gates.
Indeed, we can rewrite (\ref{U2}) as
\[
U = \left( \begin{array}{cc} 0 & 1 \\ 1 & 0  \end{array} \right)
    \left( \begin{array}{cc} 1 & 0 \\ 0 & a  \end{array} \right)
    \left( \begin{array}{cc} 0 & 1 \\ 1 & 0  \end{array} \right)
    \left( \begin{array}{cc} 1 & 0 \\ 0 & b  \end{array} \right)
    \frac{1}{2}
    \left( \begin{array}{cc} 1 + c & 1 - c \\ 1 - c & 1 + c  \end{array} \right)
    \left( \begin{array}{cc} 1 & 0 \\ 0 & d  \end{array} \right) \ ,
\]
i.e.\ a cascade of three {\tt PHASOR} gates and three {\tt NEGATOR} gates.
Two of the latter are simply {\tt NOT} gates.
In particular for the Hadamard gate, we have
\[
\hspace*{-16mm}
H = \left( \begin{array}{cc} 0 & 1 \\ 1 & 0                  \end{array} \right)
    \left( \begin{array}{cc} 1 & 0 \\ 0 & (1-i)/\sqrt{2}     \end{array} \right)
    \left( \begin{array}{cc} 0 & 1 \\ 1 & 0                  \end{array} \right)
    \left( \begin{array}{cc} 1 & 0 \\ 0 & (1+i)/\sqrt{2}     \end{array} \right)
    \frac{1}{2}
    \left( \begin{array}{cc} 1 + i & 1 - i \\ 1 - i & 1 + i  \end{array} \right)
    \left( \begin{array}{cc} 1 & 0 \\ 0 & i                  \end{array} \right) \ .
\]
Among the $3h+3g$ {\tt NEGATOR}s,
$2h+2g$ are {\tt NOT}s and
$h$ are square roots of the {\tt NOT}.

\section{Group structure}

We note that the U($n$) matrices with all line sums equal to~1
form the subgroup XU($n$) of U($n$).
For even~$n$, the XU($n$) matrices of the particular block type
\begin{equation}
    \frac{1}{2}\
    \left( \begin{array}{cc} I+V & I-V \\
                             I-V & I+V \end{array} \right) \ ,
\label{I+C}
\end{equation}
with $V \in \mbox{U}(n/2)$, form a 
subgroup\footnote{We use bXU and bZU as short notations for  
                  the block-structured XU~matrices and
                  the block-structured ZU~matrices, respectively.}
bXU($n$) of XU($n$):
\[
\mbox{U}(n) \supset \mbox{XU}(n) \supset \mbox{bXU}(n) \ ,
\]
with respective dimensions
\[
n^2 > (n-1)^2 \ge n^2/4 \ .
\]
The group structure of bXU($n$) follows directly
from the group structure of the constituent unitary matrix:
\[
     \frac{1}{2}\
    \left( \begin{array}{cc} I+V_1 & I-V_1 \\
                             I-V_1 & I+V_1 \end{array} \right) \ 
    \frac{1}{2}\
    \left( \begin{array}{cc} I+V_2 & I-V_2 \\
                             I-V_2 & I+V_2 \end{array} \right) \ =
   \frac{1}{2}\
    \left( \begin{array}{cc} I+V_1V_2 & I-V_1V_2 \\
                             I-V_1V_2 & I+V_1V_2 \end{array} \right) \ ,
\]
thus demonstrating the isomorphism bXU($n$) $\cong$ U($n/2$).

We note that the diagonal U($n$) matrices with upper-left entry equal to~1
form the subgroup ZU($n$) of U($n$).
For even~$n$, the  U($n$) matrices of the particular block type
\[
    \left( \begin{array}{cc} I &   \\
                               & V \end{array} \right) \ ,
\]
with $V \in \mbox{U}(n/2)$,
form a group bZU($n$), also a subgroup of U($n$). 
The group structure of bZU($n$) thus follows trivially from
the group structure of U($n/2$).
Whereas 
bXU($n$) is         a subgroup                  of XU($n$),
bZU($n$) is neither a subgroup nor a supergroup of ZU($n$).
Whereas $\mbox{dim(bXU(}n)) \le \mbox{dim(XU(}n))$,
the dimension of bZU($n$), i.e.\ $n^2/4$, is greater than or equal to
the dimension of  ZU($n$), i.e.\ $n-1$.

It has been demonstrated \cite{acm} that
the closure of XU($n$) and ZU($n$) is the whole group U($n$).
In other words, any member of U($n$)
can be written as a product of XU matrices and ZU matrices.
Provided $n$ is even, a similar property holds for the block versions of XU and ZU:
the closure of bXU($n$) and bZU($n$) is the whole group U($n$).
Indeed, with the help of the identity
\[
\left( \begin{array}{cc} A &   \\   & B \end{array} \right) =
\left( \begin{array}{cc}   & I \\ I &   \end{array} \right)
\left( \begin{array}{cc} I &   \\   & A \end{array} \right)
\left( \begin{array}{cc}   & I \\ I &   \end{array} \right) 
\left( \begin{array}{cc} I &   \\   & B \end{array} \right) \ ,
\]
we can transform the decomposition (\ref{F&R}) into a product containing
exclusively bXU and bZU matrices, with (among others) the
particular bXU matrix {\tiny $\left( \begin{array}{cc} & I \\ I & \end{array} \right)$},
i.e.\  the block {\tt NOT} gate.

\section{Dual decomposition}

Let $U$ be an arbitrary member of U($n$).
We apply the F\"uhr--Rzeszotnik theorem
not to $U$ but instead to its Fourier--Hadamard conjugate
$u=FUF$:
\[
u = \left( \begin{array}{cc} a & \\ & b \end{array} \right) \ F
    \left( \begin{array}{cc} I & \\ & c \end{array} \right) \ F
    \left( \begin{array}{cc} I & \\ & d \end{array} \right) \ .
\]
We decompose the left factor and insert
the $FF$ product, equal to the $n \times n$ unit matrix
{\tiny $\left( \begin{array}{cc} I & \\ & I \end{array} \right)$}:
\[
U = FuF =
    F\
    \left( \begin{array}{cc} I & \\ & ba^{-1} \end{array} \right) \ FF\
    \left( \begin{array}{cc} a & \\ &  a      \end{array} \right) \ F
    \left( \begin{array}{cc} I & \\ &  c      \end{array} \right) \ F
    \left( \begin{array}{cc} I & \\ &  d      \end{array} \right) \ F\ .
\]
Because {\tiny $F\ \left( \begin{array}{cc} a & \\ &  a \end{array} \right) \ F =
                   \left( \begin{array}{cc} a & \\ &  a \end{array} \right)$},
we obtain:
\[
U = F\
    \left( \begin{array}{cc} I & \\ & ba^{-1} \end{array} \right) \ F
    \left( \begin{array}{cc} a & \\ & ac      \end{array} \right) \ F
    \left( \begin{array}{cc} I & \\ & d       \end{array} \right) \ F\ ,
\]
a decomposition of the form
\[
U =
    \frac{1}{2}\
    \left( \begin{array}{cc} I+A' & I-A' \\ I-A' & I+A' \end{array} \right) \
    \left( \begin{array}{cc} B'   &      \\      & C'   \end{array} \right) \
    \frac{1}{2}\
    \left( \begin{array}{cc} I+D' & I-D' \\ I-D' & I+D' \end{array} \right) \ ,
\]
with 
\begin{equation}
A'=ba^{-1}, \ B'=a, \ C'=ac, \mbox{ and } D'=d \ .
\label{A'}
\end{equation}
We thus obtain a
decomposition of the form bXbZbX, dual to the F\"uhr--Rzeszotnik
decomposition of the form bZbXbZ. Just like in
the bZbXbZ decomposition, the number of degrees of freedom in
the bXbZbX decomposition exactly matches the dimension~$n^2$ of the matrix~$U$.
The diagram of the dual decomposition looks like 
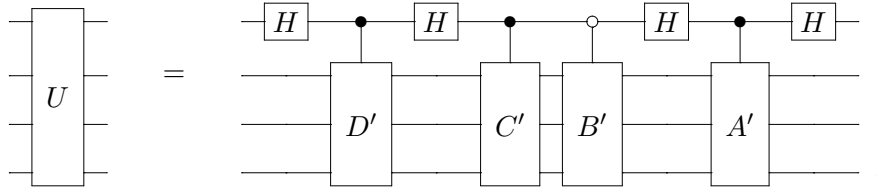
\[
\Qcircuit @C=3mm @R=3mm {
 & \multigate{3}{U} & \qw &&&   &&& & \gate{H} & \ctrl{1}          & \gate{H} & \ctrl{1}          & \ctrlo{1}         & \gate{H} & \ctrl{1}          & \gate{H}  & \qw \\
 & \ghost{U}        & \qw &&& = &&& & \qw      & \multigate{2}{D'} & \qw      & \multigate{2}{C'} & \multigate{2}{B'} & \qw      & \multigate{2}{A'} & \qw       & \qw \\
 & \ghost{U}        & \qw &&&   &&& & \qw      & \ghost{D'}        & \qw      & \ghost{C'}        & \ghost{B'}        & \qw      & \ghost{A'}        & \qw       & \qw \\
 & \ghost{U}        & \qw &&&   &&& & \qw      & \ghost{D'}        & \qw      & \ghost{C'}        & \ghost{B'}        & \qw      & \ghost{A'}        & \qw       & \qw \hspace{4mm}  \ .
}
\]

\section{Detailed procedure}

Section~2 provides the outline for the synthesis of
an arbitrary quantum circuit acting on $w$~qubits,
given its unitary transformation (i.e.\ its $2^w \times 2^w$ unitary matrix).
However, the synthesis procedure is only complete
if, given the matrix~$U$, we are able to actually compute
the four matrices $A$, $B$, $C$, and~$D$.

It is well-known that an arbitrary member~$U$ of U(2)
can be written with the help of four real parameters:
\[
U = \left( \begin{array}{rc} \cos(\fie)e^{i(\alfa+\psi)} & \sin(\fie)e^{i(\alfa+\chi)} \\
                            -\sin(\fie)e^{i(\alfa-\chi)} & \cos(\fie)e^{i(\alfa-\psi)} \end{array} \right) \ .
\]
De Vos and De Baerdemacker \cite{gent,scaling} 
noticed two different decompositions of this matrix according to (\ref{U2}):
In the former decomposition, we have
\bea 
a & = &      e^{i(\alfa+\fie+\psi)} \\
b & = &  i\, e^{i(\alfa+\fie-\chi)} \\
c & = &      e^{-2i\fie}            \\
d & = & -i\, e^{i(-\psi+\chi)} \ ,
\eea
whereas in the latter decomposition, we have
\bea 
a & = &      e^{i(\alfa-\fie+\psi)} \\
b & = & -i\, e^{i(\alfa-\fie-\chi)} \\
c & = &      e^{2i\fie}             \\
d & = &  i\, e^{i(-\psi+\chi)} \ .
\eea

F\"uhr and Rzeszotnik 
proved the generalization (\ref{F&R}) for
an arbitrary element
\[
U = \left( \begin{array}{cc} U_{11} & U_{12} \\ U_{21} & U_{22} \end{array} \right)  
\]
of U($n$), for even $n$~values,
by introducing
for each of the four $n/2 \times n/2$ matrix blocks
$U_{11}$, $U_{12}$, $U_{21}$, and $U_{22}$ of~$U$,
the polar decomposition
\[
U_{jk} = P_{jk}V_{jk}\ ,
\] 
where $P_{jk}$ is a positive-semidefinite Hermitian matrix
and   $V_{jk}$ is a unitary matrix.
Close inspection of
the proof by F\"uhr and Rzeszotnik 
(i.e.\ the proof to Theorem~8.1 in \cite{fuhr})
reveals the following expressions:
\begin{eqnarray} 
A & = &   (P_{11} + i\ P_{12})V_{11}                  \nonumber \\
B & = &   (P_{21} - i\ P_{22})V_{21}                  \nonumber \\
C & = &  V_{11}^{\dagger}(P_{11} - i\ P_{12})^2V_{11} \nonumber \\
  & = &  V_{21}^{\dagger}(P_{22} - i\ P_{21})^2V_{21} \nonumber \\
D & = & - i\, V_{11}^{\dagger}V_{12}                  \nonumber \\
  & = &   i\, V_{21}^{\dagger}V_{22} \ . \label{ABCCDD}
\end{eqnarray}
The equality of the two expressions for~$C$,
as well as      the two expressions for~$D$,
are demonstrated in the Appendix.
One can verify that
$AA^{\dagger} = BB^{\dagger} = CC^{\dagger} = DD^{\dagger} = I$,
such that $A$, $B$, $C$, and $D$ are all unitary.
For this purpose, it is necessary to observe that
$P_{11}$ and $P_{12}$ commute, as well as 
$P_{21}$ and $P_{22}$ \cite{fuhr}.
Finally, one may check that
\bea
A(I+C)  & = & 2\, U_{11} \\
B(I-C)  & = & 2\, U_{21} \\
A(I-C)D & = & 2\, U_{12} \\
B(I+C)D & = & 2\, U_{22} \ ,
\eea
such that (\ref{F&R}) is fulfilled.

It is noteworthy that there exist two formal
expressions for~$C$ and~$D$.
Whenever the polar decompositions are unique,
the two expressions evaluate to the same matrices.
However, if one~$U_{jk}$ happens to be singular,
its polar decomposition is not unique.
In this case, it is important to choose~$C$ and~$D$ consistently,
i.e.\ to take the first or second expression for both $C$ and $D$ in eqn~(\ref{ABCCDD}).

The reader will easily verify that the above 
expressions for the matrices $A$, $B$, $C$, and~$D$, for $n=2$, recover the former  
formulae    for the  scalars $a$, $b$, $c$, and~$d$.
Just like there are two different expansions in the case $n=2$,
there also exists a second decomposition in the case of arbitrary even~$n$.
It satisfies
\bea 
A & = &   (P_{11} - i\ P_{12})V_{11}  \\
B & = &   (P_{21} + i\ P_{22})V_{21} \\
C & = &  V_{11}^{\dagger}(P_{11} + i\ P_{12})^2V_{11} \\
  & = &  V_{21}^{\dagger}(P_{22} + i\ P_{21})^2V_{21} \\
D & = &   i\, V_{11}^{\dagger}V_{12} \\
  & = & - i\, V_{21}^{\dagger}V_{22} \ .
\eea

We now investigate in more detail the dual decomposition of Section~4.
Because we have two matrix sets $\{a, b, c, d \}$,
we obtain two sets $\{ A', B', C', D'\}$~:
\bea 
A' & = &   (Q_{21} - i\ Q_{22})W_{21}W_{11}^{\dagger}(Q_{11} - i\ Q_{12}) \\
B' & = &   (Q_{11} + i\ Q_{12})W_{11} \\
C' & = &   (Q_{11} - i\ Q_{12})W_{11} \\
D' & = & - i\, W_{11}^{\dagger}W_{12}
\eea
and
\bea 
A' & = &   (Q_{21} + i\ Q_{22})W_{21}W_{11}^{\dagger}(Q_{11} + i\ Q_{12}) \\
B' & = &   (Q_{11} - i\ Q_{12})W_{11} \\
C' & = &   (Q_{11} + i\ Q_{12})W_{11} \\
D' & = &   i\, W_{11}^{\dagger}W_{12} \ ,
\eea
respectively.
Here, $Q_{jk}W_{jk}$ are the polar decompositions of the four blocks~$u_{jk}$
constituting the matrix $u=FUF$.

\section{Examples}

As an example, we synthesize here the two-qubit circuit realizing the unitary transformation
\[
\frac{1}{12}\
\left( \begin{array}{cccc} 8    & 0    &  4+8i &  0    \\
                           2+ i & 3-9i &   -2i & -3-6i \\
                           1-7i & 6    & -6+2i & -3+3i \\
                           3+4i & 3-3i &  2-4i &    9i \end{array} \right) \ .
\]
We perform the algorithm of Section~5,
applying Heron's iterative method for constructing the four polar decompositions \cite{higham},
although other algorithms can be used equally.
Using ten iterations for each Heron decomposition,
we thus obtain the following two numerical results:
\[
A = \left( \begin{array}{rr}   0.67 + 0.72i      &  -0.19 + 0.03i  \\
                               0.18 + 0.06i      &   0.80 - 0.57i  \end{array} \right) , \ 
B = \left( \begin{array}{rr}  -0.33 - 0.64i      &   0.50 - 0.47i   \\
                               0.69 + 0.00i      &  -0.20 - 0.70i   \end{array} \right) , \ 
\]
\[
C = \left( \begin{array}{rr}  -0.04 - 0.95i      &  -0.01 - 0.30i   \\
                              -0.07 + 0.29i      &   0.25 - 0.92i   \end{array} \right) , \mbox{ and }
D = \left( \begin{array}{rr}   0.87 - 0.43i      &  -0.15 + 0.20i   \\
                              -0.08 - 0.24i      &  -0.68 - 0.68i   \end{array} \right)
\]
and
\[
A = \left( \begin{array}{rr}   0.67 - 0.72i      &   0.19 - 0.03i  \\
                               0.16 + 0.10i      &  -0.30 - 0.93i  \end{array} \right) , \ 
B = \left( \begin{array}{rr}   0.50 - 0.52i      &   0.50 + 0.47i   \\
                              -0.19 + 0.66i      &   0.70 + 0.20i   \end{array} \right) , \ 
\]
\[
C = \left( \begin{array}{rr}  -0.04 + 0.95i      &  -0.07 - 0.29i   \\
                              -0.01 + 0.30i      &   0.25 + 0.92i   \end{array} \right) , \mbox{ and }
D = \left( \begin{array}{rr}  -0.87 + 0.43i      &   0.15 - 0.20i   \\
                               0.08 + 0.24i      &   0.68 + 0.68i   \end{array} \right) \ .
\]

In contrast to the numerical approach in the first example, we will now perform
an analytic decomposition of a second example:
\[
U = \left( \begin{array}{rrrr} 1 &          &         &   \\
                                 &  \cos(t) & \sin(t) &   \\
                                 & -\sin(t) & \cos(t) &   \\
                                 &          &         & 1 \end{array} \right) \ ,
\]
i.e.\ a typical evolution matrix for spin-spin interaction, often discussed in physics.
We have the following four matrix blocks and their polar 
decompositions\footnote{In fact, the presented polar decompositions are only valid 
                        if $0 \le t \le \pi/2$
                        (i.e.\ if both $c \ge 0$ and $s \ge 0$).
                        However, the reader can easily treat the three other cases.}:
\bea
U_{11} & = & \left( \begin{array}{cc} 1 &  0 \\ 0 & c \end{array} \right) =
             \left( \begin{array}{cc} 1 &  0 \\ 0 & c \end{array} \right)
             \left( \begin{array}{cc} 1 &  0 \\ 0 & 1 \end{array} \right) \\
U_{12} & = & \left( \begin{array}{cc} 0 &  0 \\ s & 0 \end{array} \right) =
             \left( \begin{array}{cc} 0 &  0 \\ 0 & s \end{array} \right)
             \left( \begin{array}{cc} 0 &  y \\ 1 & 0 \end{array} \right) \\
U_{21} & = & \left( \begin{array}{cc} 0 & -s \\ 0 & 0 \end{array} \right) =
             \left( \begin{array}{cc} s &  0 \\ 0 & 0 \end{array} \right)
             \left( \begin{array}{cc} 0 & -1 \\ z & 0 \end{array} \right) \\
U_{22} & = & \left( \begin{array}{cc} c &  0 \\ 0 & 1 \end{array} \right) =
             \left( \begin{array}{cc} c &  0 \\ 0 & 1 \end{array} \right)
             \left( \begin{array}{cc} 1 &  0 \\ 0 & 1 \end{array} \right)  \ ,
\eea
where $c$ and $s$ are short-hand notations for $\cos(t)$ and $\sin(t)$, respectively.
Two blocks, i.e.\ $U_{12}$ and $U_{21}$, are singular
and therefore have a polar decomposition which is not unique:
both $y$ and $z$ are arbitrary numbers on the unit circle in the complex plane.
By choosing consistently the `second expressions' of~$C$ and~$D$, 
we find the following decompositions of~$U$:
\[
\left( \begin{array}{cccc} 1 & & & \\ & e & & \\ & & & ie \\ & & -iz & \end{array} \right) \ \frac{1}{2} \
\left( \begin{array}{cccc} 2 & & & \\ & 1+1/e^2 & & 1-1/e^2 \\ & & 2 & \\ & 1-1/e^2 & & 1+1/e^2 \end{array} \right) \
\left( \begin{array}{cccc} 1 & & & \\ & 1 & & \\ & & & -i/z \\ & & -i & \end{array} \right)
\]
and
\[
\left( \begin{array}{cccc} 1 & & & \\ & 1/e & & \\ & & & -i/e \\ & & iz & \end{array} \right) \ \frac{1}{2} \
\left( \begin{array}{cccc} 2 & & & \\ & 1+e^2 & & 1-e^2 \\ & & 2 & \\ & 1-e^2 & & 1+e^2 \end{array} \right) \
\left( \begin{array}{cccc} 1 & & & \\ & 1 & & \\ & & & i/z \\ & & i & \end{array} \right) \ ,
\]
where $e$ is a short-hand notation for $c + is$.
In spite of the singular nature of both $P_{12}$ and $P_{21}$,
this leaves only a 1-dimensional infinitum of decompositions.
The fact that some matrices~$U$ have an infinity of decompositions is further discussed in next section.

As a third and final example, we consider for~$U$ a permutation matrix.
Such choice is particularly interesting,
as a $2^w \times 2^w$ permutation matrix represents a classical reversible computation on $w$~bits
\cite{devos,wille}.
For $w=2$, we investigate the example
\[
U = \left( \begin{array}{cccc} 0 & 1 & 0 & 0 \\
                               0 & 0 & 0 & 1 \\
                               1 & 0 & 0 & 0 \\
                               0 & 0 & 1 & 0 \end{array} \right) \ .
\]
We have
\bea
U_{11} & = & \left( \begin{array}{cc} 0 &  1 \\ 0 & 0 \end{array} \right) =
             \left( \begin{array}{cc} 1 &  0 \\ 0 & 0 \end{array} \right)
             \left( \begin{array}{cc} 0 &  1 \\ x & 0 \end{array} \right) \\
U_{12} & = & \left( \begin{array}{cc} 0 &  0 \\ 0 & 1 \end{array} \right) =
             \left( \begin{array}{cc} 0 &  0 \\ 0 & 1 \end{array} \right)
             \left( \begin{array}{cc} y &  0 \\ 0 & 1 \end{array} \right) \\
U_{21} & = & \left( \begin{array}{cc} 1 &  0 \\ 0 & 0 \end{array} \right) =
             \left( \begin{array}{cc} 1 &  0 \\ 0 & 0 \end{array} \right)
             \left( \begin{array}{cc} 1 &  0 \\ 0 & z \end{array} \right) \\
U_{22} & = & \left( \begin{array}{cc} 0 &  0 \\ 1 & 0 \end{array} \right) =
             \left( \begin{array}{cc} 0 &  0 \\ 0 & 1 \end{array} \right)
             \left( \begin{array}{cc} 0 &  w \\ 1 & 0 \end{array} \right)  \ ,
\eea
where $x$, $y$, $z$, and $w$ are arbitrary unit-modulus numbers.
If, in particular, we choose $x=w=-i$ and $y=z=i$,
then we find a bZUbXUbZU decomposition of~$U$
consisting exclusively of permutation matrices:
\[
   \left( \begin{array}{cccc} 0 & 1 &   &   \\
                              1 & 0 &   &   \\
                                &   & 1 & 0 \\
                                &   & 0 & 1 \end{array} \right)
   \left( \begin{array}{cccc} 0 &   & 1 &   \\
                                & 1 &   & 0 \\
                              1 &   & 0 &   \\
                                & 0 &   & 1 \end{array} \right)
   \left( \begin{array}{cccc} 1 & 0 &   &   \\
                              0 & 1 &   &   \\
                                &   & 0 & 1 \\
                                &   & 1 & 0 \end{array} \right) \ .
\]
In the next section,
we will demonstrate that such is possible 
for any $n \times n$ permutation matrix
(provided $n$ is even).

\section{Light matrices and classical computing}

The second and third example in previous section
lead us to a deeper analysis of sparse unitary matrices. 
                             
{\bf Definition:}
Let $M$ be an $m \times m$ matrix
with, in each line and each column, maximum one non-zero entry.
We call such sparse matrix `light'.
Let $\mu$ be the number of non-zero entries of~$M$.
We call $\mu$ the weight of~$M$. We have $0 \le \mu \le m$.
If $\mu = m$, then $M$ is  regular;
if $\mu < m$, then $M$ is singular.
The reader will easily prove the following two lemmas:

{\bf Lemma 1:}
Let $PU$
(with $P$ a positive-semidefinite matrix and $U$ a unitary matrix)
be the polar decomposition of a light matrix~$M$.
Then $P$ is a diagonal matrix and $U$ is a complex permutation matrix.
If $\mu$, the weight of~$M$, equals $m$, then $U$ is unique;
otherwise, we have an $(m-\mu)$-dimensional infinity of choices for~$U$.

{\bf Lemma 2:}
If $P$ is a diagonal matrix and $U$ is a complex permutation matrix,
then $U^{\dagger}PU$ is a diagonal matrix, with the same entries as $P$,
in a permuted order.

We now combine these two lemmas.
Assume that the $n \times n$ matrix $U$ consists of four $n/2 \times n/2$ blocks,
such that the two blocks $U_{11}$ and $U_{12}$ are light.
Then, by virtue of Lemma~1, the positive-semidefinite
matrices $P_{11}$ and $P_{12}$ are diagonal.
Therefore $P_{11}-iP_{12}$ is diagonal and so is $(P_{11}-iP_{12})^2$.
By virtue of Lemma~1 again, the matrix $V_{11}$ is a complex permutation matrix.
Finally, because of Lemma~2,
the matrix $C=V_{11}^{\dagger}(P_{11}-iP_{12})^2V_{11}$ is diagonal
and so are $I+C$ and $I-C$.
As a result, for $n=2^w$, the matrix
$F {\tiny \left( \begin{array}{cc} I   &     \\     & C   \end{array} \right)} F = \frac{1}{2} \
   {\tiny \left( \begin{array}{cc} I+C & I-C \\ I-C & I+C \end{array} \right)}$
represents a cascade of $2^{w-1}$ {\tt NEGATOR} gates acting on the first qubit
and controlled by the $w-1$ other qubits:
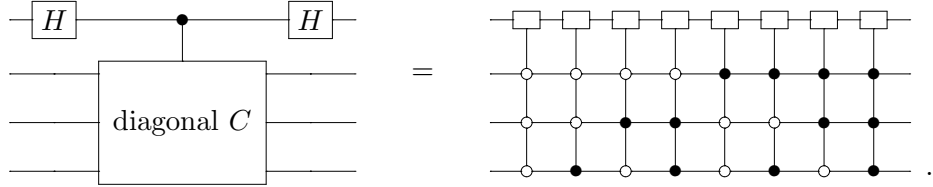
\[
\Qcircuit @C=3mm @R=3mm {
 & \gate{H} & \ctrl{1}                         & \gate{H} & \qw  &&&   &&& & \gate{\ }  & \gate{\ }  & \gate{\ }  & \gate{\ }  &  \gate{\ }  & \gate{\ }  & \gate{\ }  & \gate{\ } & \qw \\
 & \qw      & \multigate{2}{\mbox{diagonal }C} & \qw      & \qw  &&& = &&& & \ctrlo{-1} & \ctrlo{-1} & \ctrlo{-1} & \ctrlo{-1} &  \ctrl{-1}  & \ctrl{-1}  & \ctrl{-1}  & \ctrl{-1} & \qw \\
 & \qw      &        \ghost{\mbox{diagonal }C} & \qw      & \qw  &&&   &&& & \ctrlo{-1} & \ctrlo{-1} & \ctrl{-1}  & \ctrl{-1}  &  \ctrlo{-1} & \ctrlo{-1} & \ctrl{-1}  & \ctrl{-1} & \qw \\
 & \qw      &        \ghost{\mbox{diagonal }C} & \qw      & \qw  &&&   &&& & \ctrlo{-1} & \ctrl{-1}  & \ctrlo{-1} & \ctrl{-1}  &  \ctrlo{-1} & \ctrl{-1}  & \ctrlo{-1} & \ctrl{-1} & \qw \hspace{4mm}  \ .
}
\]

We now are in a position to discuss the case of $U$ being 
an $n \times n$ permutation matrix. 
Its special interest results from the fact that, for $n$ equal to a power of~2,
such matrix represents a classical reversible computation.

First, we will prove that 
$\frac{1}{2} \ {\tiny \left( \begin{array}{cc} I+C & I-C \\ I-C & I+C \end{array} \right)}$
is a structured permutation matrix.
If $U$ is an $n \times n$ permutation matrix, 
then both $n/2 \times n/2$ blocks $U_{11}$ and $U_{12}$ are light,
the sum of their weights $\mu_{11}$ and $\mu_{12}$ being equal to $n/2$.
The matrices $P_{11}$ and $P_{12}$ are diagonal, with
entries equal to~0 or~1,
with the special feature that, wherever there is a zero entry in $P_{11}$,
the matrix $P_{12}$ has a~1 on the same row, and vice versa.
The matrix $P_{11}-iP_{12}$ thus is diagonal, with
all diagonal entries either equal to~1 or to~$-i$.
Hence, the matrix $(P_{11}-iP_{12})^2$ is diagonal, with
all diagonal entries either equal to~1 or to~$-1$,
and so is matrix~$C$.
Hence, the matrices $I+C$ and $I-C$ are diagonal with entries either~0 or~2.
As a result, for $n=2^w$, the matrix
$F {\tiny \left( \begin{array}{cc} I   &     \\     & C   \end{array} \right)} F = \frac{1}{2} \
   {\tiny \left( \begin{array}{cc} I+C & I-C \\ I-C & I+C \end{array} \right)}$
represents a cascade of 1-qubit {\tt IDENTITY} and {\tt NOT} gates acting on the first qubit
and controlled by the $w-1$ other qubits.
Thus the above $2^{w-1}$ {\tt NEGATOR} gates 
all equal a classical gate: 
either an {\tt IDENTITY} gate or a {\tt NOT} gate.

Next, we proceed with proving that $D$ is also a permutation matrix.
The matrices $V_{11}$ and $V_{12}$ are 
complex permutation matrices. 
The matrix $V_{11}$ contains $n/2$ non-zero entries.
Among them, $n/2-\mu_{11}$ can be chosen arbitrarily,
$\mu_{11}$ being the weight of $U_{11}$.
We denote these arbitrary numbers by $x_j$, 
in analogy to $x$ in the third example of Section~6.
Analogously,
we denote by $y_k$ the $n/2-\mu_{12}$ arbitrary entries of $V_{12}$.
Because $U$ is a permutation matrix,
the weight sum $\mu_{11}+\mu_{12}$ necessarily equals $n/2$.
The matrix $-iV_{11}^{\dagger}V_{12}$ also is a complex permutation matrix
and thus has $n/2$ non-zero entries.
This number matches the total number of degrees of freedom
$(n/2-\mu_{11}) + (n/2-\mu_{12}) = n/2$.
Because $U$ is a permutation matrix, $V_{11}$ and $V_{12}$
can be chosen such that the non-zero entries of the product
$-iV_{11}^{\dagger}V_{12}$ depend only on an $x_j$ or on an $y_k$ 
but not on both.
More particularly these entries are 
either of the form $-i/x_j$ 
    or of the form $-iy_k$. 
By choosing 
all $x_j$ equal to~$-i$ and 
all $y_k$ equal to~$i$,
the matrix $-iV_{11}^{\dagger}V_{12}$, and thus $D$, is a permutation matrix. 

Because $U$, 
$\frac{1}{2} \ {\tiny \left( \begin{array}{cc} I+C & I-C \\ I-C & I+C \end{array} \right)}$, and 
${\tiny \left( \begin{array}{cc} I & \\ & D \end{array} \right)}$ are permutation matrices, also
${\tiny \left( \begin{array}{cc} A & \\ & B \end{array} \right)}$ is an $n \times n$ permutation matrix.
Ergo: given an $n \times n$ permutation matrix~$U$,
we can construct four $n/2 \times n/2$ permutation matrices
$A$, $B$, $C$, and~$D$.
Threfore, we recover here the Birkhoff decomposition method for permutation matrices
and thus, for $n=2^w$, a well-known synthesis method for classical reversible logic circuits
\cite{devos,storme,vanrentergem},
based on the Young subgroups of the symmetric group {\bf S}$_{2^w}$

\section{Conclusion}

Thanks to the F\"uhr and Rzeszotnik decomposition of U($n$) matrices with even~$n$,
and three more decompositions presented above,
we can synthesize the quantum circuit
performing an arbitrary unitary transformation from U($2^w$),
in four systematic and straightforward ways.
The present bZbXbZ  and
            bXbZbX decompositions are more practical
than the       ZXZ decomposition because
no Fourier transforms $F_j$ (with $2 \le j \le 2^w$) are necessary. Only
controlled XU(2) or {\tt NEGATOR}s and
controlled ZU(2) or {\tt  PHASOR}s are necessary.
Alternatively, one can apply
controlled {\tt  PHASOR}s combined with
controlled Hadamard gates, i.e.\ $F_2$ transforms.

In contrast to previously developed synthesis methods for quantum circuits
(based e.g.\  on the sine-cosine or the KAK or the Householder decomposition),
the present four matrix decompositions naturally include
the synthesis of classical reversible circuits.

\appendix

\section*{Appendix}

{\bf Lemma 3:} 
Let $P$ and $P'$ be positive-semidefinite matrices,
let $U$ and $U'$ be           unitary matrices, and
let $PU=P'U'$.
Then, $U$ is equal to~$U'$, provided $P$ and $P'$ are regular.
 
{\bf Lemma 4:} 
Let $P_j$ and $U_j$ be positive-semidefinite and unitary matrices, respectively.
Then any equality of the form
$P_1U_1P_2U_2P_3U_3... = P_1'U_1'P_2'U_2'P_3'U_3'...$ implies
$   U_1   U_2   U_3... =     U_1'    U_2'    U_3'...$,
provided all $P_j$ and all $P_j'$ are regular.
The proof is based on repeated application of 
$PU=UQ$ with $Q=U^{\dagger}PU$ also a positive-semidefinite matrix,
followed by use of Lemma~3.~\ $\blacksquare$

From the unitarity condition
$U^{\dagger}U = UU^{\dagger} = \left( \begin{array}{cc} I & 0 \\ 0 & I  \end{array} \right)$ 
follows:
\begin{eqnarray}
P_{11}^2 + P_{12}^2 & = & I      \nonumber \\
P_{21}^2 + P_{22}^2 & = & I      \nonumber \\
V_{11}^{\dagger}P_{11}^2V_{11} + V_{21}^{\dagger}P_{21}^2V_{21} & = & I \label{1} \\
V_{12}^{\dagger}P_{12}^2V_{12} + V_{22}^{\dagger}P_{22}^2V_{22} & = & I \label{2} \ ,
\end{eqnarray}
as well as
\begin{eqnarray}
P_{11}V_{11}V_{21}^{\dagger}P_{21} + P_{12}V_{12}V_{22}^{\dagger}P_{22} & = & 0 \nonumber \\
V_{11}^{\dagger}P_{11}P_{12}V_{12} + V_{21}^{\dagger}P_{21}P_{22}V_{22} & = & 0 \label{3} \ .
\end{eqnarray}
If $P_{11}$, $P_{12}$, $P_{21}$, and $P_{22}$ are regular, then,
by virtue of Lemma~4, this leads to
\begin{eqnarray}
V_{11}V_{21}^{\dagger} & = & - V_{12}V_{22}^{\dagger} \label{4} \\
V_{11}^{\dagger}V_{12} & = & - V_{21}^{\dagger}V_{22} \label{5} \ .
\end{eqnarray}

In the expression 
\[
V_{11}^{\dagger}(P_{11} - i\ P_{12})^2V_{11} 
\]
or 
\[
  V_{11}^{\dagger}P_{11}^2V_{11}
-iV_{11}^{\dagger}P_{11}P_{12}V_{11}
-iV_{11}^{\dagger}P_{12}P_{11}V_{11}
- V_{11}^{\dagger}P_{12}^2V_{11} \ ,
\]
we eliminate 
$P_{11}^2$     with the help of (\ref{1}),
$P_{11}P_{12}$ with the help of (\ref{3}),
$P_{12}P_{11}$ with the help of (\ref{3}), and
$P_{12}^2$     with the help of (\ref{2}).
Subsequently, we eliminate $V_{11}$ and 
$V_{11}^{\dagger}$ with the help of (\ref{4}-\ref{5}).
We thus obtain
\bea
& &
  V_{21}^{\dagger}P_{22}^2V_{21}
-iV_{21}^{\dagger}P_{21}P_{22}V_{21}
-iV_{21}^{\dagger}P_{22}P_{21}V_{21}
- V_{21}^{\dagger}P_{21}^2V_{21} 
\\ & = &
  V_{21}^{\dagger}(P_{22}-iP_{21})^2V_{21} \ .
\eea

\section*{Acknowledgement}

The authors thank the European COST Action IC~1405 `Reversible computation'
for its valuable support.

\end{document}